# Demystifying BERT: Implications for Accelerator Design

Suchita Pati[1], Shaizeen Aga[2], Nuwan Jayasena[2], Matthew D. Sinclair[1,2]

[1]University of Wisconsin-Madison
{spati,sinclair}@cs.wisc.edu

[2]Advanced Micro Devices Inc.
{shaizeen.aga,nuwan.jayasena}@amd.com

**Abstract**

*Transfer learning in natural language processing (NLP), as realized using models like BERT (Bi-directional Encoder Representations from Transformer), has significantly improved language representation with models that can tackle challenging language problems. Consequently, these applications are driving the requirements of future systems. Thus, we focus on BERT, one of the most popular NLP transfer learning algorithms, to identify how BERT's algorithmic behavior can guide future accelerator design. To this end, we carefully profile BERT training and identify key algorithmic behaviors which are worthy of attention in accelerator design.*

*We observe that while computations which manifest as matrix multiplication dominate BERT's overall runtime, as in many convolutional neural networks, memory-intensive computations also feature prominently. We characterize these computations, which have received little attention so far. Further, we also identify heterogeneity in compute-intensive BERT computations and discuss software and possible hardware mechanisms to further optimize these computations. Finally, we discuss implications of these behaviors as networks get larger and use distributed training environments, and how techniques such as micro-batching and mixed precision training scale. Overall, our analysis identifies holistic solutions to optimize systems for BERT-like models.*

## 1. Introduction

In recent years, rapid advancements in natural language processing (NLP) have enabled important applications that interpret human language, such as intelligent personal assistants, near instantaneous language translation, and more intelligent search engines. This tremendous, transformative rise has been enabled by a virtuous synergy of (1) better hardware systems, (2) larger datasets, and (3) improved deep neural network (DNN) structures and machine learning (ML) algorithms that further benefit from more efficient hardware and larger datasets. For decades, technology scaling has driven Moore's Law and enabled more efficient hardware designs. However, a key challenge to more fully realizing ML's potential is that further progress on designing more efficient hardware is hampered by the slowing of technology scaling and Moore's Law [41]. In recognition of the slowing of technology scaling, the community responded by optimizing ML workloads for GPUs [15, 20, 42, 57, 89] as well as designing accelerators like FPGAs [23] and TPUs [40, 46, 55, 80]. As a result, ML workloads, especially convolutional neural networks (CNNs) and ML inference, are often able to fully utilize on-chip ALUs and execute efficiently [1, 2, 4, 12, 16, 23, 25, 26, 30, 31, 37, 38, 40, 47, 48, 52, 53, 62, 57, 65, 67, 72, 77, 84, 85, 87, 88].

However, transformer-based networks, which fine-tune massive, pre-trained models for specific language-related tasks, have emerged as a key trend in applying ML to NLP tasks. These models, like **B**i-directional **E**ncoder **R**epresentation from **T**ransformer **(BERT)** [19], mark a major shift in this trend towards deeper knowledge transfer by applying entire models to different tasks. Since its introduction, BERT has produced state-of-the-art results for several sentence-level and word-level NLP tasks [19], which has further made BERT the basis of several larger (3.9 billion parameters) and more tuned models [68, 51]. Moreover, BERT has also inspired many other recent NLP architectures and NLP models such as ALBERT [49], ERNIE2.0 [74], Google's TransformerXL [18], OpenAI's GPT-2 [59] and GPT-3 [14], RoBERTa [51], and XLNet [81]. This explosion in ML algorithms designed for NLP highlights the importance of further optimizing future systems for these algorithms [83]. Although each of these models are worthy of deeper investigation, in this paper we focus on BERT, since it embodies several of the key trends that are important when designing and optimizing future accelerators for Transformer-based networks.

BERT has become the state-of-the-art model for language representation because it overcomes the shortcomings of its predecessors to accurately train on billions of words from unlabeled datasets. For example, BERT's transformer-based architecture uses deeply bi-directional contextual understanding, implying its representation of a token in a given sequence encodes information from all the preceding and subsequent tokens. As a result, BERT models are large, consisting of 110-340 million parameters, which helps make them extremely accurate. Consequently, computations such as gradient descent optimizers [64] that update model parameters can feature much more prominently in both computation and memory requirements for BERT as compared to other ML models like CNNs and older RNN-based NLP models. Moreover, BERT uses an attention-based structure that avoids sequential bottlenecks [76] (discussed further in Section 2). Collectively, these features enable BERT to effectively handle longer sentences and allow training models on datasets which were previously considered too large, both of which are key contributors to the power of its model and accuracy.

| Num | Takeaway | Algorithmic Explanation | Section |
|---|---|---|---|
| 1 | Of the different layers in BERT, the transformer layers dominate its training time, while the output & embedding layers have negligible contribution. | BERT is predominantly made up of transformer layers which scale with deeper models. | 3.2.1 |
| 2 | BERT's gradient descent optimizer (LAMB), which updates the model weights, is the second highest contributor to BERT's training runtime, and its contribution increases with decreasing input token count per iteration. | LAMB is a complex optimizer algorithm designed for large (effective) batch-size convergence to update 340 million parameters in the BERT (Large) model. | 3.2.1 |
| 3 | Weight updates, using LAMB, are more important to optimize for with mixed-precision training. | LAMB updates are computed using single precision copies of parameters and gradients, while rest of the training computations use half precision data. | 3.2.1 |
| 4 | GEMMs from the linear transform and fully connected sub-layers dominate the transformer layer's execution time, while the remainder of the execution time is spent executing several smaller non-GEMM operations. | GEMMs in these layers operate on both input, activations, and weights, and have a computational complexity of $O(n^3)$ while other operations predominantly process activations with a complexity of $O(n^2)$. | 3.2.1 |
| 5 | Non-GEMM and LAMB update operations become more important to optimize with reduced precision training. | GEMMs speedup much more than other operations in half precision due to faster arithmetic and reduced memory traffic. | 3.2.1 |
| 6 | Unlike RNNs, a mini-batch size of one does not lead to matrix-vector operations in BERT. | GEMM dimensions in BERT are a multiple of the input token count (mini-batch size x sequence length) and layer's hidden dimension and scale with these parameters. | 3.2.2 |
| 7 | Not all GEMMs in BERT are equal: only some of them can fully utilize highly parallel accelerators. | GEMMs in an attention sub-layer are smaller or skinnier (compared to that of a FC sub-layer) and may also be memory-bound. | 3.2.2 |
| 8 | Parameter updates using LAMB are extremely memory intensive. | LAMB reads and writes 4× more data (gradients, weights, and optimizer states) than the model size while executing few elementwise operations. | 3.2.3 |
| 9 | Non-GEMM phases that make up to 30-40% of BERT Large's (FP32) training time are memory-bound. | These phases consist of series of elementwise (add, mul, scale) and reduction operations. | 3.2.3 |
| 10 | Optimizing memory-bound operations is even more important for BERT with reduced precision training. | Memory-bound elementwise & reduction operations benefit from a reduced footprint, but not much from half precision arithmetic (unlike GEMMs). | 3.2.3 |
| 11 | Decreasing input mini-batch size or sequence length, i.e., the total number of tokens per iteration, increases the proportion of memory-intensive LAMB. | LAMB operates on weights, gradients, and optimizer parameters which are independent of input token count. | 3.3.1 |
| 12 | Both transformer and LAMB parameter update remain important as transformer layer count is increased. | Operations in all but the input and output layers scale linearly with transformer layer count. | 3.3.2 |
| 13 | The runtime proportion of GEMMs and LAMB update increase in wider models (larger hidden dimensions). | The number of parameters (inputs to both GEMMs and LAMB) scale quadratically with transformer hidden dimension. | 3.3.2 |
| 14 | The per-device compute and memory-bound operation breakdown in data-parallel multi-device training is similar to that of single-device training. | The model is replicated on each device and the majority of the communication (of gradients) is overlapped with computation. | 4.1.2 |
| 15 | The proportion of memory-bound LAMB update operations decreases with model-parallel multi-device training. Its communication volume and time are larger than that with data-parallel training and increases with increased model parallelism. | Model parameters are divided across multiple devices. Communication of activations (& error in backprop) is serialized with computations and its volume (and time) increases as batch size is scaled (to compensate for smaller operations). | 4.1.2 |

Table 1: Summary of key takeaways

Understanding the underlying behaviors of these features is vital to designing efficient accelerators for BERT and similar models. For example, BERT's transformer layers intersperse compute-intensive and memory-intensive (e.g., low op-to-byte ratio) operations and exhibit heterogeneity in both compute-intensive and data-intensive operations (discussed further in Section 3 and 5). However, there has been little work so far in their detailed characterization. Most prior works on ML algorithms focus on CNNs, recurrent neural networks (RNNs), or recommendation models [2, 28, 80, 88]. Moreover, prior works on analyzing Transformers fall short of the details required to build efficient systems for them. Consequently, recent works aimed at building accelerators for BERT contain matrix-vector engines for layers which actually perform matrix-matrix operations [29, 32]. Thus, in this work we provide a detailed algorithmic characterization of BERT's operations, their manifestation, size, and nature, as well as their contribution to the overall execution runtime. Table 1 summarizes the key takeaways. Overall, we closely investigate BERT and identify BERT's algorithmic behavior to guide future accelerator design, making the following key contributions:

- We carefully characterize BERT training and identify key algorithmic behaviors which future accelerator designs should optimize for.
- We observe that, similar to many CNNs, compute-intensive (e.g., matrix multiplications) operations make up the majority of BERT's overall runtime. However, unlike most CNNs, data-intensive computations also feature prominently in BERT (34% of execution time), highlighting the importance of designing systems that provide high performance for both compute- and data-intensive tasks.
- We further characterize these data-intensive computations in BERT and identify significant heterogeneity in these computations, which require different optimizations in future systems.
- We also identify inherent heterogeneity in well-studied compute-intensive operations in BERT and discuss software and possible future hardware mechanisms to optimize them.
- Finally, we discuss the implications of these behaviors as networks get larger and are trained in a distributed training environment, and how they interact with techniques like micro-batching and mixed precision training.

More broadly, our analysis identifies holistic future acceleration solutions for Transformer-based models like BERT.

## 2. Background

### 2.1. Transfer Learning & BERT Pre-training

Transfer learning is a general ML principle where a model trained for a particular task is reused as a starting point for a different task or the knowledge gained in learning a task is transferred while training for a different task. Although transfer learning has been widely used in Computer Vision [66, 82], it was only recently applied to NLP in BERT. As shown in Figure 1, BERT's training model consists of two parts: a long pre-training phase (Figure 1a) where the model learns the language, independent of any of the target tasks, and a short fine-tuning phase (Figure 1b) where the pre-trained model is tuned further for a specific task (it can be tuned separately for several different tasks once pre-trained). For pre-training, the model is trained on a large unlabeled dataset (e.g., Wikipedia and the BookCorpus), and it learns the language via the two unsupervised tasks: masked word prediction (Masked-LM) and Next Sentence Prediction (NSP). Although there are 24 different versions of BERT publicly available [75], we focus on the largest and most common version: (*BERT Large*), which uses additional layers to improve accuracy (discussed further in Section 2.3). BERT Large's pre-training can take up to four days on 16 cloud TPU Pods, where each Pod contains four TPU chips and 64 GB RAM [27].

BERT's pre-training is further split into two phases. In Phase-1 the model is trained on sequences of length 128 and in Phase-2 it is trained on sequences of length 512. To reduce training time 90% of pre-training iterations are in Phase-1 and only 10% of those are in Phase-2: although longer sequences are important for improved model accuracy, BERT's runtime is quadratic with the input sequence length (discussed further in Section 2.2). Finally, during fine-tuning (Figure 1b), the pre-trained BERT model is trained on a labeled dataset for a specific task with minimal changes to the model architecture. Fine-tuning is usually inexpensive, taking up to an hour on a single TPU Pod [27] or few hours on a GPU, and BERT can be fine-tuned for 11 different tasks with state-of-the-art accuracy.

### 2.2. Attention

The attention network within the transformer encoder layers, as shown in Figure 2(c, d), is a key component of BERT and other recent NLP models. Given an input sequence, the attention networks output a representation of the sequence such that each output token within the sequence is encoded with *weighted* information from all other tokens in the sequence. In particular, each output token is a *weighted* sum of all *value* vectors and these *weights* are calculated for every pair of *query* and *key* vectors in the input sequence. The query, key and value vectors of every token are the input token vector itself in case it has a single attention head.

However, there are usually multiple attention heads (16 in BERT Large) in each attention layer, where each attention head calculates an independent set of attention weights, thus capturing additional properties and relationships of/between the tokens. To do so, the input vectors of all the tokens in the sequence are first linearly projected into three independent sets of vectors (for query, key, and value) such that every element in the output vectors contains information about every element in their respective input vector. Each of these vectors are then split across all the attention heads and used to calculate the aforementioned weighted sum. The outputs from the heads are later concatenated and projected to obtain the final output. Thus, since every output token is obtained independently using all input tokens and the attention scores used are also obtained for every token pair independently, the attention layer completely removes the sequential dependencies of its predecessor, RNNs. Furthermore, since many of these components are all-to-all, attention layer computations grow quadratically with the input sequence length, making it computationally expensive for longer sequence-length inputs.

### 2.3. BERT Network Model

BERT's basic building block is the transformer encoder layer; multiple transformer encoder layers are shown stacked in Figure 2(a). A transformer encoder layer, as shown in Figure 2(b), consists of an attention layer and a fully connected (FC) feed-forward layer, both of which are followed by a residual connection and layer normalization. BERT also has an input embedding layer that provides the transformer layers with an input representation for every token in the input sequence. It is constructed by summing their corresponding token, segment, and position embeddings. Finally, BERT has an output classification layer responsible for two tasks: Masked-LM and NSP. An additional output layer may be added later during fine-tuning based on the target task the model is used for. Currently, BERT has two widely used model configurations: BERT Large, which consists of 24 transformer encoder layers and has larger internal parameters, and BERT Base, a smaller version of the model. This work focuses on the larger and more accurate configuration of BERT.

### 2.4. Gradient Descent Optimizer & LAMB

*Gradient descent* is the most common algorithm used to train neural networks. It works by minimizing an *objective function* (usually the loss) parameterized by the model's parameters. It does so by updating the parameters in the opposite direction of the gradient of the objective function with respect to the parameters. It uses a *learning rate* to determine the size of the step it takes per iteration in that direction. Models and frameworks today use various algorithms to further optimize gradient descent to converge faster. These optimizers help derive appropriate learning rates for different parameters in the model (which is useful for sparse models) as well as at dif-ferent stages of training. These optimizers require computing and tracking additional parameters per iteration besides the model weights and gradients, such as *momentum* and *velocity*, which requires additional memory capacity.

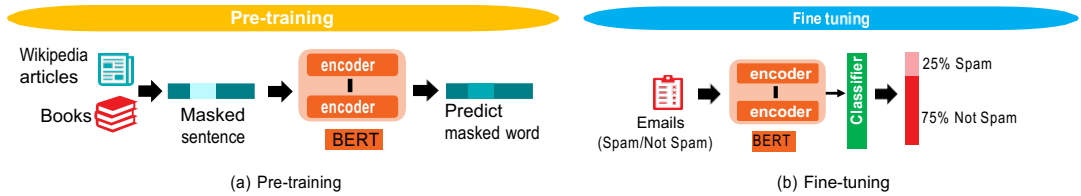

(a) Pre-training

(b) Fine-tuning

**Figure 1: BERT Training overview.**

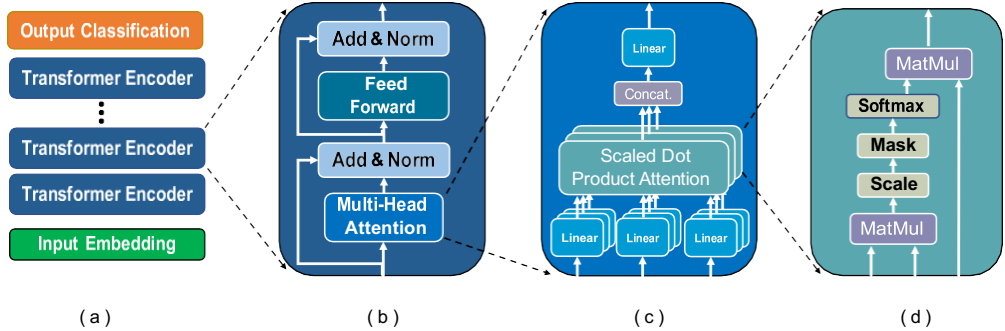

**Figure 2: BERT hierarchical model breakdown.**

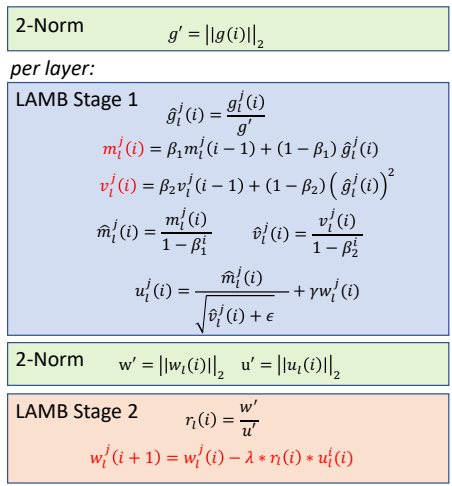

**Figure 3: LAMB Algorithm**

Although BERT is compatible with many different optimizers, recently BERT has been used with the LAMB optimizer, which has proven effective when used with very large batch-sizes and distributed training [83]. Figure 3 details the LAMB algorithm, which updates the model parameters at the end of the model's forward and backward gradient calculations and is executed in two stages; the first determines the update values and learning rate multiplier using additional momentum (m) and velocity (v) parameters from the past iterations and gradients of the current iteration. In the second stage, the model weights are updated with the output of the first stage. This pair of two stages is executed independently for every layer in the model, with each set accessing an independent set of data (weights, gradients and optimizer parameters of the corresponding layer). LAMB is executed once every few, if gradient accumulation is enabled – discussed further in Section 4.2) iteration(s).

### 2.5. Distributed Training

Growing model sizes and datasets have made the use of multiple devices to train DNNs commonplace. Large models with millions or billions of parameters have large memory capacity requirements often not available in a single device. Furthermore, as datasets grow, the number of training iterations required to converge (along with better accuracy) also grow. Oftentimes, systems also use multiple devices working in parallel to improve scalability.

**Data Parallelism**: The most common and straightforward technique is to use *data parallelism*. In data parallel training, the model is replicated on every device and the input dataset is partitioned amongst the $D$ devices. Each device then iterates over its own dataset (using a batch of $b$ samples every iteration) and trains its model while synchronizing with all the other devices every iteration.[1] During synchronization, the local gradients from all the devices are averaged and re-distributed using an AllReduce operation and each model then updates its set of parameters using these accumulated gradients. This data parallelism enables very large mini-batch size training which otherwise would not be feasible with a single device's memory capacity.

**Model Parallelism**: The other approach used in distributed training is *model parallelism* in which the model is split across multiple devices, with each device responsible for a disjoint portion of the model. This enables training of deeper and/or larger models as each device now only has to store a subset of the parameters and activations in its memory. Model parallelism can be further classified as *inter-layer* and *intra-layer*; the former splits the model layer-wise with every device responsible for a (set of contiguous) layer(s) and the latter splits individual layers amongst multiple devices. Both require communication of activations (in the forward pass) as well as input gradients or errors (in the backward pass). The type and size of computations as well as communication differs based on the specific implementation.

---
[1]This does not hold for asynchronous training.

**Hybrid Parallelism**: Models today also take a ***hybrid*** approach, where the model is split between $M$ devices in a cluster (model parallel), and there are $D$ such clusters, each of which trains on a disjoint set of the input dataset (data parallel). Overall, this enables training of a model using $M*D$ devices.

## 2.6. Arithmetic Intensity

The arithmetic intensity of an operation is defined as the number of computations it performs for every byte of data it reads. It is equivalent to its compute intensiveness and helps determine if the operation will be limited by memory. If the operation performs very few computations on each byte of data read, it is most likely going to be bottlenecked by the memory latency of the system and vice-versa. It is an important parameter used while designing accelerators, for example, high arithmetic intensity operations would benefit from more compute, whereas low arithmetic intensity operation require higher memory bandwidth or larger on-chip memory to avoid frequent memory accesses.

## 3. BERT's Algorithmic Behavior

In this section, we first describe the system setup we use to study BERT, following which we provide a detailed runtime breakdown of BERT's execution and highlight key operations to optimize for. And finally, we describe these operations in detail, their manifestation, sizes, and nature to further guide system designers.

### 3.1. Experimental Setup

**3.1.1. System** Many accelerators, including GPUs and TPUs are being used to train BERT. In this study we choose to study BERT on a GPU because of its wide availability as well as its popularity for DNN training. However, our key takeaways are accelerator agnostic and should be applicable to other accelerators (discussed further in Section 5). We study BERT pre-training on a system consisting of an AMD Ryzen™ Threadripper™ CPU [7] and an AMD Instinct™ MI100 GPU [10] with 32GB HBM2 [39]. Our software stack is built on top of the AMD ROCm™ platform [8] with PyTorch v1.7. Since our goal is to analyze and characterize BERT training in a platform independent manner, we focus on relative importance of operations in BERT, as well as the size and nature of operations. Our takeaways mainly depend on the BERT's network architecture, its hyperparameters, and the training mechanisms used, and thus are largely unaffected by the choice of hardware and software stack.

**3.1.2. BERT Phases** During the pre-training phase, BERT's model is trained on large unlabeled datasets. Since this phase is the more expensive part of training BERT, usually taking several days, we focus on analyzing it rather than the fine-tuning phase. Moreover, execution of the fine-tuning phase that follows pre-training is largely the same, requiring only minor tweaks to the model. Therefore, studying pre-training provides a good understanding of BERT's fine-tuning behavior and focuses on the costliest – and thus most important to accelerate – part. We focus on the characteristics of Phase-1 of pre-training in which the model is trained on inputs of sequence length ($n$) 128 with a mini-batch size ($B$) of 32, although we also discuss how these characteristics differ in Phase-2 when $n$ = 512. Finally, we study both single as well as mixed precision training and further discuss how bottlenecks shift with reduced precision.

**3.1.3. BERT Hyperparameters** We execute BERT pre-training using the BERT Large configuration and the English Wikipedia corpus [19]. The BERT Large model consists of 24 transformer layers ($N$) with a hidden state size ($d_{model}$) of 1024, 16 attention heads ($h$) and an intermediate dimension ($d_{ff}$, usually $4*d_{model}$) of 4096. Note however that since these hyperparameters can scale in future models, it is also important to understand their impact on BERT's execution profile which we describe in Section 3.3. Throughout the remainder of the paper, we use these symbols to refer to the parameters, as shown in Table 2.

**3.1.4. Profiling Mechanism** Given the enormous corpus of data BERT is trained on, pre-training it can span several days making it impractical to profile its entire execution. Prior work on characterizing CNNs usually approximate the execution by profiling a single training iteration [88] whereas recent works have shown that NLP model iterations can be heterogeneous due to varying input sequence length [58]. However, unlike other NLP models which use labeled datasets, BERT is trained over unlabeled datasets derived from plain text, providing users the ability to generate fixed-length input sequences. Therefore, within a phase, training iterations in BERT operate on same-size inputs, implying the iterations remain largely homogeneous in terms of computations. Thus, we profile and study a single training iteration (after a set of warm-up iterations) per pre-training phase which is representative of the entire training execution. We use rocProf [6], a performance analysis tool, to gather runtime and other performance counter data.

| Acronym | Parameter |
|---------|-----------|
| B | mini-batch size |
| $d_{model}$ | Hidden Dimension |
| h | #Attention Heads |
| $d_{ff}$ | Intermediate Dim. |
| N | Layer Count |
| n | Input Sequence Length |

Table 2: BERT Parameters

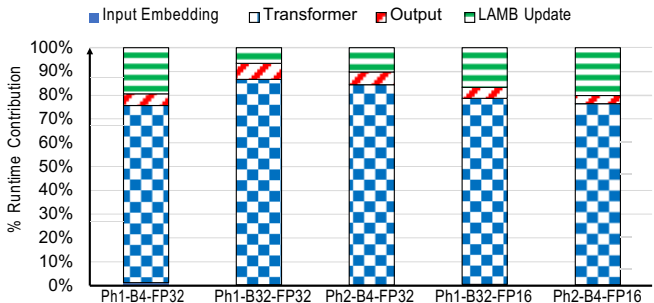

Figure 4: Runtime Breakdown of BERT pre-training for different phases, mini-batch sizes, and precisions.

### 3.2. Compute & Memory Demands of BERT Operations

Next we study the computations in an end-to-end training iteration of BERT which entails (a) a forward phase in which the model processes input sequences and produces an output, (b) a back-propagation phase where the network calculates the loss in output prediction and calculates weight gradients, and (c) update phase where the weights are updated based on the gradients. Table 2 maps the abbreviations to their corresponding BERT hyperparameters, and Table 3 describes the BERT GEMM operations and activation sizes. For every important BERT layer, Table 3 lists three GEMM operations: for the forward (FWD) phase, for the gradient in activation (or error) calculation, and for the gradient in weight calculation. The latter two are executed during backpropagation (backprop, or BWD). Note that, in our discussions throughout the paper, we represent a matrix as $M$x$N$, a GEMM as $M$x$N$x$K$, and a product of two variables as $a*b$.

**3.2.1. Runtime breakdown** We first present a high-level runtime breakdown amongst different network layers and training phases. In all the runtime distribution plots in the paper, we consider a layer's forward and back-propagation phase together and show the weight updates separately. Figure 4 shows this breakdown of BERT pre-training for different phases, mini-batch sizes, and precisions: Ph$i$-B$j$-FP$k$, where $i$ represents the BERT pre-training phase (Phase-1 or Phase-2), $j$ represents the mini-batch size, and $k$ represents the precision (i.e., number of bits) used in the experiment.

For all the different configurations, the transformer layers dominate the iteration runtime, which is intuitive given they account for most of the layers. The output layer constitutes only a small proportion and the time taken by the input embedding layer is negligible. Interestingly, the LAMB optimizer (Section 2.4) is consistently the second highest contributor. The proportion of LAMB optimizer increases as the number of tokens ($n*B$) per training iteration decrease (e.g., Ph1-B4-FP32 and Ph2-B4-FP32 have a much higher proportion of LAMB than Ph1-B32-FP32). This happens because the runtime of forward propagation and backward gradient calculations are dependent on the token count, while the weight update runtime is only proportional to the model size (and independent of the input token count).

The proportion of LAMB updates increase with mixed-precision (MP) training (in Ph1-B32-FP16 and Ph2-B4-FP16) as well. In MP training, inputs and model weights are converted to half precision during forward and backward propagation of the model and the gradients calculated are then used to update a master copy of weights in single precision. Reduced precision helps speed up computations via faster arithmetic as well as reduced memory reads/writes. Therefore, we observe that both the forward and backprop operations speed up by about 2X while the runtime of LAMB updates remains constant.

*Key Takeaway 1*: Transformer layer runtime dominate BERT training time and therefore, they should be optimized for. The output and embedding layers have negligible contribution to the runtime.
*Key Takeaway 2*: LAMB parameter updates are the second highest contributor to BERT's training runtime. Furthermore, its runtime contribution increases with decreasing token count per iteration.
*Key Takeaway 3*: LAMB updates become more important to optimize for with mixed-precision training.

Figure 5 presents a hierarchical breakdown of transformer layers (with labels representing their contribution to the overall training time) from Figure 2. The second bar, **Transformer**, shows the runtime breakdown of the components of the transformer layer, i.e., the attention layer, the Fully Connected (FC) feed-forward layer, as well as the dropout (DR), residual (Res), and Layer Normalization (LN) layers combined. Overall, the FC layer has a higher runtime contribution compared to the attention layer because the intermediate dimension of the FC layers ( Section 3.1.3) is much larger (4X). Additionally, the combined DR, Res and LN layers have a smaller, but non-negligible (7% in FP32, 9% in MP) contribution to the iteration runtime.

The third bar of Figure 5, **Attention**, shows that a significant portion of the runtime (22% in FP32, 19% in MP) is spent on linear transform operations. These operations are required to project the inputs query, key, and value vectors (of length $d_{model}$) into $h$ different, learned linear projections (of dimension $d_{model}/h$) to be operated on by the $h$ attention heads. The batched-GEMM (details in Section 3.2.2) operations corresponding to calculation of attention scores and weighted average of tokens constitute a much smaller proportion (7% in FP32, 10% in MP) of the overall runtime. The feed-forward sub-layer of BERT's transformer layers (last bar in Figure 4) consist of two **fully connected (FC)** connections with a GeLu (Gaussian Error Linear Unit) [34] activation in between. The FC layers dominate the runtime with the GeLu activation contributing to a non-negligible 14% (in FP32, 10% in MP).

| Operation | FWD | | | | BWD Grad. Activation | | | | BWD Grad. Weight | | | |
|---|---|---|---|---|---|---|---|---|---|---|---|---|
| | M | N | K | Batch | M | N | K | Batch | M | N | K | Batch |
| Linear Trans. | $d_{model}$ | $n*B$ | $d_{model}$ | - | $d_{model}$ | $n*B$ | $d_{model}$ | - | $d_{model}$ | $d_{model}$ | $n*B$ | - |
| Attn. Score | $n$ | $n$ | $d_{model}/h$ | $B*h$ | $n$ | $d_{model}/h$ | $n$ | $B*h$ | $d_{model}/h$ | $n$ | $n$ | $B*h$ |
| Attn. O/p | $d_{model}/h$ | $n$ | $n$ | $B*h$ | $d_{model}/h$ | $n$ | $n$ | $B*h$ | $n$ | $n$ | $d_{model}/h$ | $B*h$ |
| FC-1 | $d_{ff}$ | $n*B$ | $d_{model}$ | - | $d_{model}$ | $n*B$ | $d_{ff}$ | - | $d_{model}$ | $d_{ff}$ | $n*B$ | - |
| FC-2 | $d_{model}$ | $n*B$ | $d_{ff}$ | - | $d_{ff}$ | $n*B$ | $d_{model}$ | - | $d_{ff}$ | $d_{model}$ | $n*B$ | - |

**Table 3: Architecture-Agnostic Sizes of BERT GEMMs for both training and inference.**

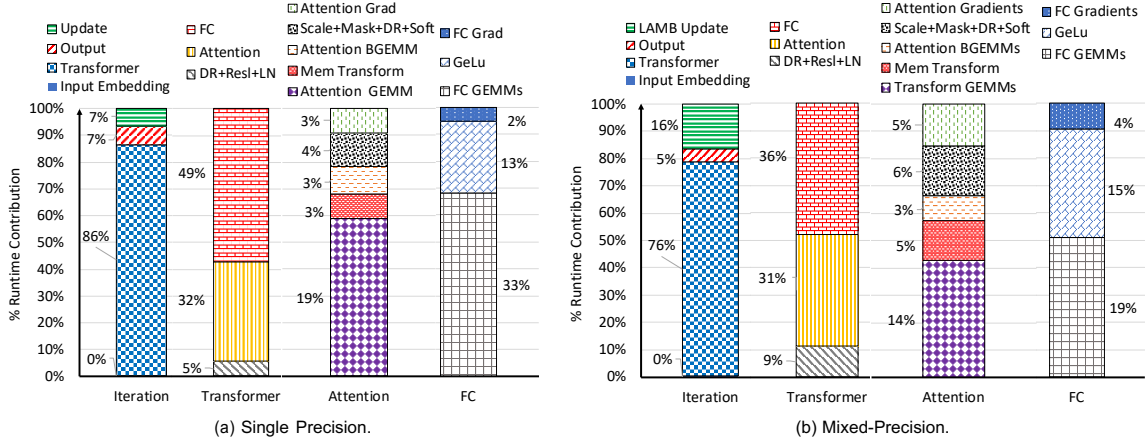

**Figure 5: Execution time breakdown of BERT's pre-training Transformer layers.**

Finally, the proportion of the linear transform and FC layers falls considerably (from 57% in FP32 to and 40% in MP) as compared to other operations when executed with reduced precision implying they benefit more from the drop in precision. These layers manifest as GEMMs (details in Section 3.2.2) and their speedup can be attributed to both faster arithmetic (Matrix Core Engine [9]) and smaller memory footprint. The smaller memory footprint in turn also leads to improved cache effectiveness thus enabling the data-reuse in GEMMs.

*Key Takeaway 4*: The linear transform and the fully connected layers dominate the transformer layer's execution, contributing to about 57% (in FP32, and 40% in mixed precision) of the BERT's iteration runtime. Rest of the time is spent executing several smaller operations discussed in detail next.

*Key Takeaway 5*: Since reducing the precision improves GEMM speedup more than other operations, non-GEMM operations are important to optimize for at reduced precision.

**3.2.2. GEMM operations in BERT** Since a significant amount (60% in FP32 and 45% in MP) of the BERT's per-iteration execution time is spent executing GEMMs, we next characterize the different types of GEMMs executed in BERT and their compute requirements. There are three sets of GEMMs in BERT's transformer layers: the ones for the attention (score and weighted average) computations, the ones for the linear transform operation which enable multiple attention heads, and the ones corresponding to the fully connected layers.

As illustrated in Figure 6 (within the dotted box), the attention head takes the query and key vectors of all the tokens in the input and calculates the attention score (compatibility) between every token pair through the product of their respective query and key vectors, executed as a GEMM. Since there are $h$ independent attention heads working in parallel, there are $h$ such GEMMs executed in parallel per input sequence. Furthermore, since each training iteration usually operates on a *mini-batch* of inputs ($B$), there are $B*h$ GEMMs invoked as a single batched-GEMM kernel (**Attention B-GEMM** in Figure 5a, 5b, and the Attn. Score in Table 3)). The attention scores/weights are then used to calculate the weighted sum of all the value vectors in the input sequence, also invoked as batched-GEMM operation (Attn. O/p in Table 3) with $B*h$ parallel GEMMs. While there are several ($B*h$) parallel GEMMs in this operation, each of these GEMMs is quite small (dimensions of $n$ and $d_{model}/h$).

To enable multiple attention heads, the query, key, and value vectors of all the tokens are first linearly projected (illustrated outside the dotted box in Figure 6) into $h$ smaller ($d_{model}/h$) feature vectors. All the token vectors of all the input sequences in the mini-batch are usually combined into a single matrix (with dimensions $B*n$ x $d_{model}$), and using the learned linear Weights ($W_k$, $W_q$, $W_v$) are projected via three different GEMMs which take up most of the attention layer runtime (**Linear Transform GEMMs** in Figure 5a, 5b and

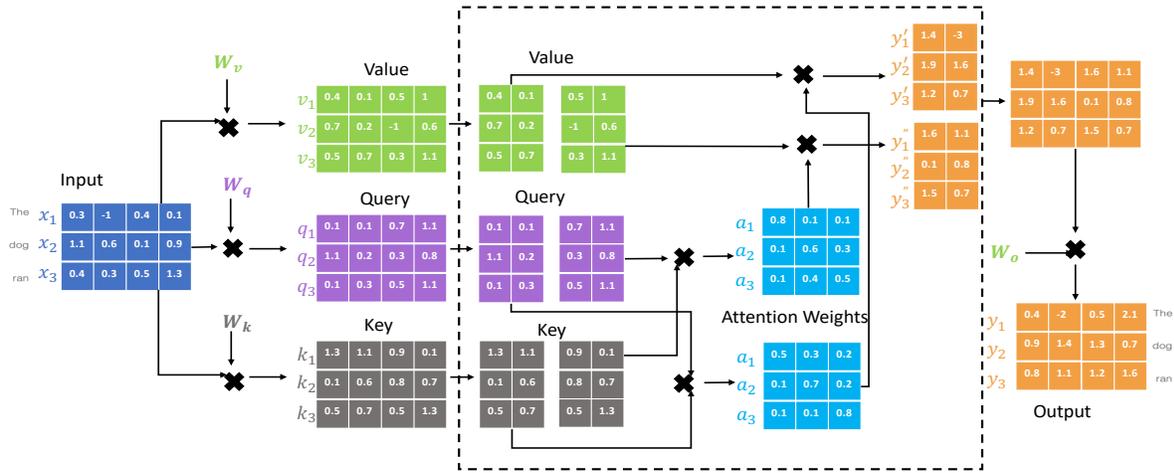

**Figure 6: Computations in the Attention Layer**

Table 3). The output of these GEMMs is then split to create the query, key, and value vectors for each of the attention heads. The outputs of the attention operations are then concatenated into a single matrix which is then projected back using $W_o$ (Figure 6 right). Finally, the FC layers are processed as two large GEMMs (see Table 3 FC-1 & FC-2 for their dimensions) and dominate the execution time of the FC layer (Figures 5a and 5b).

Usually, GEMMs with larger and squarer matrices perform better on modern accelerators. The larger matrix size helps GEMMs leverage the compute power of highly parallel accelerators. Along with a squarer shape, it enables them to exhibit better data reuse, thus improving overall cache effectiveness and hiding memory access latency. However, we find that not all GEMMs and B-GEMMs in BERT have such compute requirements. To understand better we plot the ops-per-byte ratio of all the GEMMs (labeled as *transposeA, transposeB, M, N, K, [batch]*) that appear in a BERT's transformer layer (Ph1 with $B = 32$ and FP32 precision) in Figure 7. While we find that FC GEMMs are large and extremely compute intensive, the attention's linear transform GEMMs are not, with 4X smaller matrix dimensions and smaller ops-per-byte ratios. Furthermore, the matrices in the B-GEMM operations of the attention layers are even smaller, leading to extremely low ops/byte ratio. These GEMMs have much higher memory bandwidth requirements (Figure 8) compared to the other two sets of GEMMs, making them memory-intensive in contrast to the popular belief of GEMMs being compute intensive.

*Key Takeaway 6*: GEMM dimensions in BERT are a multiple of the input token count (i.e., the product of input mini-batch and sequence length or $B*n$), and layer's hidden dimension ($d_{model}$ or $d_{ff}$) and scale with these parameters. Unlike RNNs, a mini-batch size of one does not lead to matrix-vector operations.

*Key Takeaway 7*: Not all GEMMs in BERT are equal. Smaller, skinnier GEMMs in BERT's attention layer are memory-

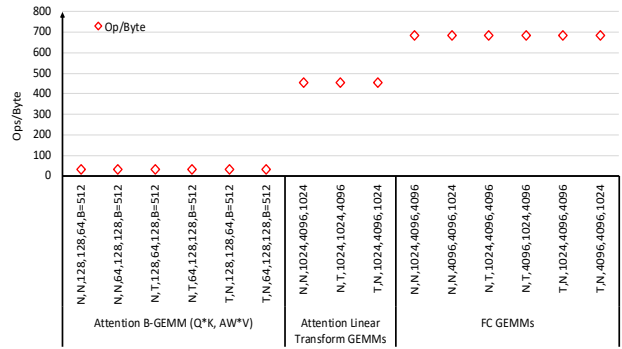

**Figure 7: Measuring the arithmetic intensity of BERT's training GEMMs. The results show that not all of BERT's GEMMs are equal.**

bound and under-utilize highly parallel accelerators.

**3.2.3. Non-GEMM operations in BERT** In the previous sections, we observed that 40% (FP32) to 55% (MP) of BERT's training time is spent executing non-GEMM operations. Thus, along with GEMMs, these operations are key to building efficient accelerators for BERT-like models. Figure 8 plots the memory bandwidth requirements of these operations normalized to the maximum bandwidth achieved by any BERT operation (i.e., the element-wise multiplication operation). The figure also shows the theoretical operations-per-byte ratio (or arithmetic intensity) of these operations, a metric to indicate the inherent nature of these algorithms. We also include these metrics for key GEMM kernels of the transformer layer for comparison. There are four key parts of BERT pre-training where we observe these operations: (1) LAMB, (2) attention head, (3) GeLU activation, and (4) dropout, residual connection, and layer normalization.

**LAMB Updates**: As detailed in Figure 3, *LAMB* is responsible for updating the model parameters and is executed in two stages; the first (LAMB Stage 1 in Figure 8) determines the update values and learning rate multiplier using additional momentum (m) and velocity (v) parameters from the past iterations and gradients of the current iteration (all of which are the same size as that of the parameters, i.e., $M$x$N$ of the BWD Grad. Weight GEMMs in Table 3). This stage performs multiple elementwise (EW) add, multiply, divide, scale, and square-root operations on these parameters and therefore, has very low ops/byte ratio (Figure 8) making it memory intensive. In the second stage (LAMB Stage 2 in Figure 8), the model weights are updated with the output of stage 1 using multiple EW operations and therefore LAMB stage 2 has similar memory characteristics to LAMB stage 1. This set of two stages (kernels) is executed once per layer in the model, with each set accessing an independent set of data (weights, gradients, and optimizer parameters of different layers). Therefore, each set has no data reuse across kernels (we discuss fusion's impact in Section 5.1.1). Moreover, LAMB must perform the L2 Norm (reduction) across all the gradients of the model before it can update any parameter, which serializes the model update with respect to the entire model backprop. Overall, LAMB is 7-20% of an iteration's runtime, which increases even further with increasing transformer layer size in larger models, smaller token size per iteration, and MP training.

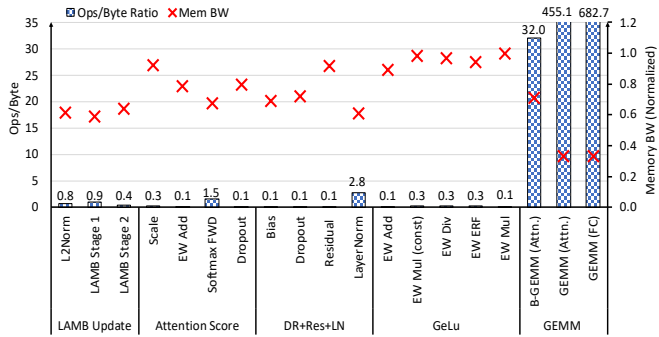

**Figure 8: Arithmetic intensity and bandwidth requirements of BERT operations.**

*Key Takeaway 8*: The LAMB optimizer is extremely memory-intensive, reading 4X more data than the model size while executing few EW operations.

**Attention Head**: *The attention head* generates the attention score (compatibility) between token pairs and uses them to calculate the weighted representation of each token in the input. The attention scores, before being used, however are normalized and apply the *softmax* activation and dropout function (Scale, Mask, DR, Soft. in Figures 5a and 5b). Like LAMB, these operations also have a poor ops/byte ratio (Figure 8), making them memory latency bound. The memory bandwidth requirements of these operations are also high, especially in the backward pass, due to larger inputs, making some of these kernels memory bandwidth bound.

**GeLU**: *GeLu activation* operation is executed between the two FC GEMMs and consists of a sequence of EW ops each with a very low ops/byte ratio, as shown in Figure 8. This along with the large input activation size (output of FC GEMM), also makes these kernels both memory bandwidth as well as latency bound.

**Dropout, residual connection, and Layer Normalization (DR+Res+LN)**: *The dropout, residual connection and the Layer Normalization (LN)* operations executed after each of the attention and FC layers as shown in Figure 8 have very low arithmetic intensity and large memory bandwidth requirements. The dropout and residual connection manifest as EW multiplication and add operations, respectively, while the LN layer is a reduction operation. While these operations are a small proportion of the overall execution time, their proportion increases with MP (Figure 5, 5).

While LAMB kernels remain unchanged in MP training (updates are in FP32), the memory bandwidth bound kernels in most of the above phases are sped up by about 1.5-1.9X in MP. This speedup, however, is much smaller than that of GEMMs, leading to an increase in proportion of these operations in MP training. Therefore, non-GEMM operations become even more important to optimize for when training with reduced precision.

*Key Takeaway 9*: BERT training has multiple memory-bound (both latency and bandwidth) elementwise operations that make up to 30-40% of its (FP32) runtime.

*Key Takeaway 10*: Optimizing memory-bound operations is even more important for BERT when reduced precision train- ing is used, since with reduced precision memory-bound operations make up 50% of all operations.

### 3.3. Effects of Hyperparameter Sweep on BERT Characteristics

As models get larger/deeper, hyperparameters like the transformer layer count, hidden dimension, attention head count and mini-batch size are bound to change. Therefore, it is important to understand how its characteristics would change as the model evolves. Accordingly, we next analyze and characterize the impact of BERT's training hyperparameters on its execution profile. We further characterize the nature of those operations and their sizes in terms of network hyperparameters, to help architects make well-informed decisions when designing such accelerators.

#### 3.3.1. Input Size: mini-batch size, input sequence length

The mini-batch ($B$) size and the input sequence length ($n$) are important hyperparameters as they can impact training convergence; increasing $B$ improves training throughput but can also lead to convergence degradation, especially in data-parallel training (Section 4.1) while a larger $n$ is always beneficial for a language model but is expensive to train on. Both together also decide the number of tokens that BERT processes in a single iteration and thus dictate the memory capacity as well as computational requirements of BERT pre-training. Increasing these hyperparameters increases the total computations

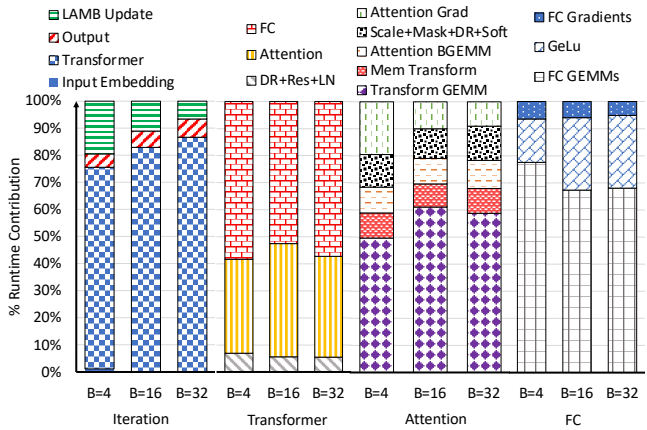

Figure 9: Impact of scaling mini-batch size

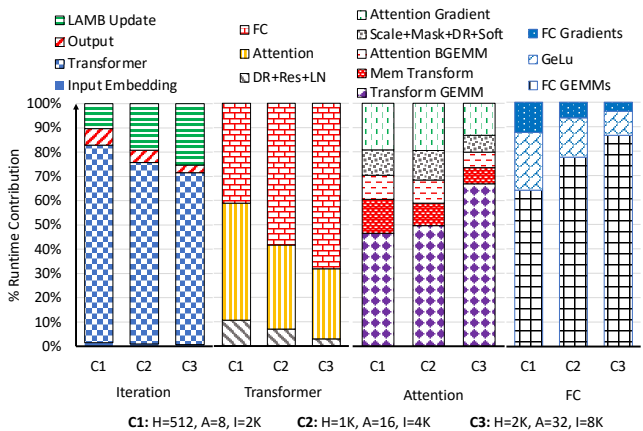

**C1**: H=512, A=8, I=2K   **C2**: H=1K, A=16, I=4K   **C3**: H=2K, A=32, I=8K

Figure 10: Impact of Scaling Transformer Layer Size

in the forward and backward gradient calculations, although the parameter update computation (which depends only on the model size) remains constant. Figure 9 highlights this: as $B$ ranges from 4 to 32, LAMB updates make up a relatively larger proportion of an iteration's runtime when batch size is small, and a smaller proportion when mini-batch size is large. The rest of the changes in breakdown can be attributed to the type (quadratic vs linear) of operations in the layers. Therefore, the proportion of most EW operations (e.g., add, mul, tanh) increases while that of GEMMs decreases as $B$ or $n$ is scaled down.

*Key Takeaway 11:* Decreasing mini-batch size ($B$) or input sequence length ($n$), i.e., the total number of tokens per iteration ($B*n$), increases the proportion of memory-intensive LAMB and other elementwise operations.

**3.3.2. Model Size: transformer layer count, hidden dim.** BERT's model size is dictated by the number of transformer layers ($N$) and their size, via the hidden and intermediate dimensions, as well as the attention head count. The increase in transformer layer count scales the count of every operation pertaining to the transformer layer (including LAMB update, since model parameters also scale). Intuitively, this does not impact the breakdown within a transformer layer. However, both the transformer layers and the LAMB update see a slight increase in their overall proportions since the number of operations within the input embedding and output layers remain constant. The increase in hidden and intermediate dimensions, on the other hand, increases the input size to the layers and operations. Figure 10 highlights this change in distribution as layer size varies. Since the number of computations in GEMMs are quadratic with the size of the input matrices, their runtime scales much more than that of other EW operations, which scale linearly with the input. Thus, the proportion of linear GEMMs in Attention and the FC GEMMs increase in proportion. However, since the FC layer is 4X the size of the attention layer, FC's proportion scales much more than the Attention layer. Finally, the proportion of LAMB increases considerably as the size of each transformer layer increases. Unlike the linear scaling of parameter count with transformer layer count, the parameter count is quadratic with the hidden dimension of layers (if $H = 1024$, parameters = 1024*1024).

*Key Takeaway 12:* Both transformer and LAMB updates scale linearly and therefore remain important as transformer layer count is increased in deeper models.

*Key Takeaway 13:* Runtime proportion of GEMMs and LAMB increase in wider models with larger transformer layers.

## 4. Effects of Training Mechanisms on BERT Characteristics

Although studying BERT training on a single device is important and reveals interesting computational behaviors, BERT is usually trained in a multi-device environment which enables training of larger, deeper models and significantly accelerates training. Furthermore, there are several techniques used to accelerate per-device training.

### 4.1. Multi-device Training

As described in Section 2, multi-device DNN training usually uses either data parallel, model parallel, or both. Data parallel training enables large (effective) batch size training, thus reducing training time, while model parallel training enables the training of deeper, wider models which cannot fit in a single device due to memory capacity requirements. Therefore, we characterize BERT using a data-parallel approach and a state-of-the-art (intra-layer) model-parallel technique, Megatron-LM, using BERT Large and 128 GPUs with 2-way model and 64-way data parallelism [68]. Since MegaTron's global mini-batch size is fixed to 1024, the per iteration mini-batch size is 16 (1024/64).

**4.1.1. Modeling Multi-device Training** We construct the per-device execution profile in a distributed setting using a single device (GPU in our case) and an analytical model, described in detail below:

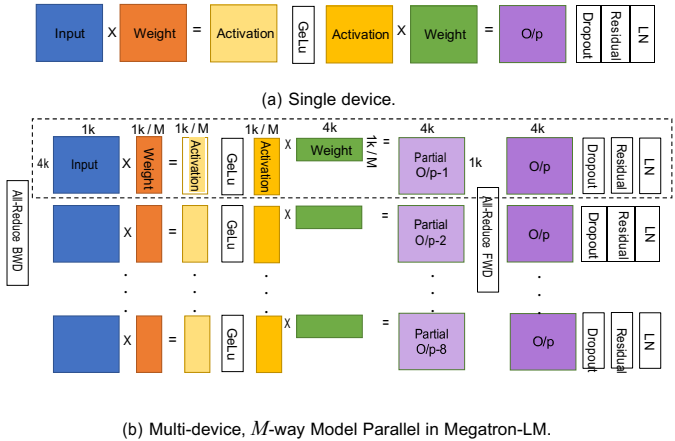

(a) Single device.

(b) Multi-device, $M$-way Model Parallel in Megatron-LM.

**Figure 11: FC Layer Computation in BERT.**

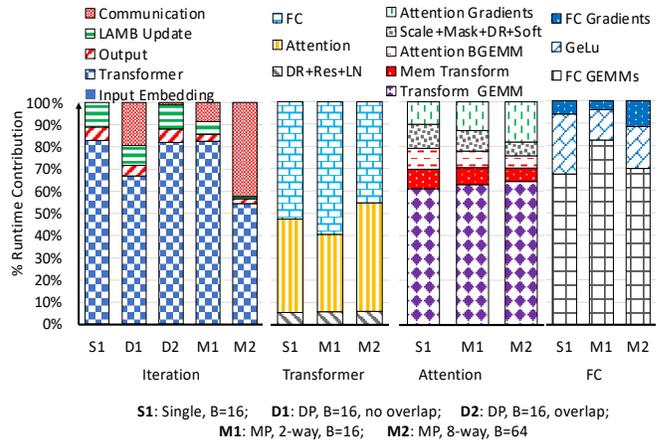

**S1**: Single, B=16;  **D1**: DP, B=16, no overlap;  **D2**: DP, B=16, overlap;
**M1**: MP, 2-way, B=16;  **M2**: MP, 8-way, B=64

**Figure 12: BERT iteration breakdown for Multi-GPU training.**

**Modeling Data Parallel Training**: Since data parallel training replicates the model on every device, the per-device computation matches single-device training. Additionally, an AllReduce operation gathers each device's gradients (during backprop). To estimate the AllReduce communication costs, we use the size of gradients and use the Ring AllReduce algorithm to model the data that will be sent and received per device. To estimate the communication time, we assume communication via PCIe™ 4.0 and divide the size of the gradients by its communication bandwidth. Finally, since the communication and computations of different layer's gradients are independent, they can be overlapped to speed up training (e.g., layer $L$'s gradients can be communicated while the device calculates gradients for layer $L-1$). We model this overlap by taking the maximum of the computation and communication times for every pair of consecutive layers.

**Modeling Model Parallel Training**: Unlike data-parallel training, Megatron-LM splits most of the layer's operations across all the devices, some of these operations (e.g., weight matrices) are split horizontally while others are split vertically. The remaining layers (e.g., LayerNorm) are replicated across devices to reduce communication overheads. Figure 11 illustrates this change for the FC layer operations. The operations within the dashed line illustrate the dimension of operations on each device after a $M$-way model-parallel split of BERT Large. To model these computations, we execute all the operations in each BERT layer with input dimensions that are expected after the splitting and replication of the layers.

The LAMB operations are also split, as each device is responsible for only a fraction of the parameters. Finally, there are four additional AllReduce operations executed per transformer layer. We estimate this communication time with the expected activation and gradient sizes using the approach described above. However, unlike in data parallel approach, these AllReduce operations are serialized with other operations and thus cannot be overlapped with computations.

**4.1.2. Multi-GPU Training Profile** Figure 12 compares the execution breakdown of different distributed training mechanisms. We also plot one single-GPU with $B = 16$ (Single, B=16) for reference.

**Data Parallel**: The per-GPU execution profiles of BERT's data parallel approach with overlap, D1 (DP, B=16 with overlap), is similar to a single GPU training, S1 (w/ B=16). This is unsurprising because the model is replicated across all the GPUs, with each model computing the forward, backprop and update phases independently. Although the data parallel model has additional inter-GPU synchronization for the local gradients, this cost (except for the first layer) is hidden by overlapping computations and using a fast channel such as PCIe 4.0™ D2 (DP, B=16, without overlap) highlights this: D2 uses the same data parallel approach as D1, but the gradients are communicated after the entire backprop completes. As a result, a significant portion of D2's runtime (19%) is spent communicating gradients since they cannot be overlapped with computation. Recent work on reducing the cost of replicated optimizer states and operations in data parallel ML training [60] could potentially reduce the proportion of LAMB in the data parallel execution profiles.

**Model Parallel**: Figure 12 shows the runtime breakdown for the model parallel, multi-GPU implementations: M1 (MP, 2-way with B=16) and M2 (MP, 8-way model parallelism with B=64). The high-level iteration breakdown M1 is similar S1, single-GPU training with the same mini-batch size (16) However, there are two key differences. First, M1 spends more of its runtime (9%) communicating both activations and gradients. Second, the proportion of LAMB scales by half since each device is responsible for half of the model's parameters. These changes are even more prominent in M2. The communication costs increase to about 42% in M2 due to the larger volume of data communicated (due to its larger $B$). The proportion of LAMB in M2 is negligible. Finally, unlike the transformer layers which process both inputs/activations and weights, LAMB processes weights (along with gradients and LAMB parameters, all of which are the same size as the weights), which scale inversely with the number of devices.

*Key Takeaway 14:* The compute and memory-bound operation breakdown in a data-parallel multi-GPU setting is similar to that of single-GPU training due to data parallel's ability to overlap most communication with computation.

*Key Takeaway 15:* The proportion of memory-bound LAMB operations drop with model parallel training since the parameter count per device scales inversely with device count. However, the communication volume (and runtime) increase with increased model parallelism due to a larger mini-batch size.

### 4.2. Micro-batching & Gradient Accumulation

Similar to data-parallelism, micro-batching increases the effective batch size of training by splitting a very large mini-batch into smaller *micro-batches*, each of which can fit within a single GPU's memory [35]. The model is then trained on a single micro-batch at a time, without updating the model parameters. Instead, the gradients calculated using all the micro-batches in a mini-batch are averaged using *gradient accumulation* and a single update is applied to the model parameters. It reduces the cost of updates by the micro-batch count but adds additional EW scale and add operations to accumulate gradients per micro-batch.

## 5. Potential Optimizations for BERT

### 5.1. Software Mechanisms

Since roughly half of BERT's training time is GEMMs, a straight-forward way to improve throughput would be to improve the overall accelerator FLOPs. However, as these GEMMs speed up, the remaining memory-intensive operations will become the bottleneck.

**5.1.1. Kernel Fusion** Fusion is an optimization in which two or more consecutive phases (e.g., kernels in GPUs) in an application, potentially with a producer-consumer relationship, are combined into a single phase. Fusing kernels can significantly improve performance by increasing data reuse, reducing duplicate memory accesses, and reducing kernel launch overhead. Thus, it is common for ML libraries to fuse phases together, especially for GPU inference, to reduce overhead [11, 21, 22, 50, 56, 71, 73, 79]. However, there are several considerations while fusing kernels:

**Data reuse across consecutive operations**: the extent of data reuse across the kernels being fused is directly correlated with improved performance. Kernel fusion leverages data reuse between operations by keeping data from being flushed into global memory between kernel calls [5, 33, 45, 69, 70]. Therefore, data intensive operations like all but the LAMB kernels in Section 3.2.3 (GeLu activation, dropout+residual+LN, and attention head operations) in which the output of one operation feeds the next are perfect scenarios for applying kernel fusion. Interestingly, the LAMB kernels are already fused in PyTorch (e.g., all the operations in Stage 1 of Figure 3 are fused into the LAMB Stage 1 kernel of Figure 8). Thus, there is little or no benefit from further fusion given the independent data (parameter, gradients, optimizer states of different layers) accessed by the remaining LAMB kernels.

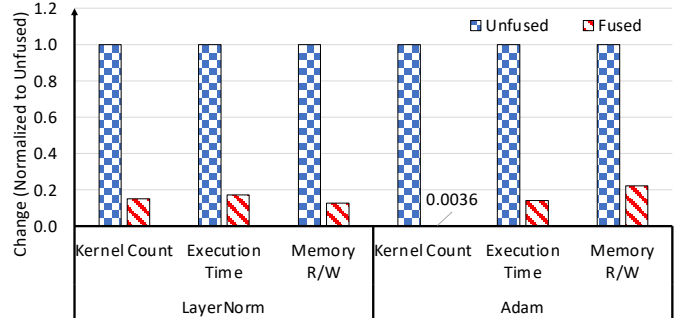

**Figure 13: Impact of fusing kernels, normalized to the unfused baseline.**

Figure 13 examines how effective fusion is at reducing kernel counts, execution time and memory accesses compared to the unfused baselines. We examine two key algorithms: the LayerNorm function and the Adam optimizer. [2] For LayerNorm, fusion is very effective, reducing kernel count, execution time, and memory traffic by 6-8X. Fusion similarly reduces Adam's kernel count drastically from thousands to tens of kernels. However, Adam's execution time and memory traffic do not scale similarly since their reduction only comes from fusion of optimizer kernels per layer and not across different layers.

**Reusability of fused kernels**: to use kernel fusion, libraries must provide fused implementations of the operations of interest, which require additional software engineering. Therefore, it is key that sequence of consecutive kernels being fused are commonly occurring across models. Amongst the three scenarios in BERT which could benefit from kernel fusion, the only scenario which will have re-usability is the GeLu activation operations. The other sets of operations, while common, are not guaranteed to appear in that order, limiting the re-usability of a fused implementation.

**5.1.2. GEMM Fusion** Another common optimization to increase parallelism within GEMMs on modern accelerators is to fuse independent smaller GEMMs, with a common input matrix, into a single large GEMM. Figure 14 shows how the independent linear transform GEMMs of the attention layers can be fused into a single operation. Since these GEMMs operate on the same input matrix and their respective weight matrices, $W_q, W_k$, and $W_v$ (Figure 14, left), they can be fused together as shown in Figure 14 (right) where the weight matrices are concatenated, and the input matrix is read once. The output of this GEMM is simply the output of the three individual GEMMs concatenated, which can be split for subsequent use. Figure 15 examines the impact of fusing these linear GEMMs in both the forward and backward pass (labeled as $M, N, K, transposeA, transposeB$). As shown, fusing these smaller GEMMs together improves performance by up to 62% by enabling better data reuse across operations (the common matrix is read only once) and increasing parallelism by using a larger matrix dimension. Its impact is higher when the input matrices are small (due to a smaller token count or hidden dimension).

---
[2]Adam is an alternate to LAMB; we chose Adam because its unfused and fused versions were publicly available.

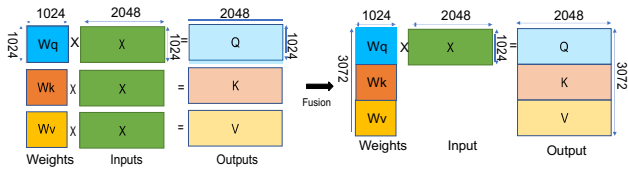

Figure 14: Fusion of Linear GEMMs in BERT's Attention layer.

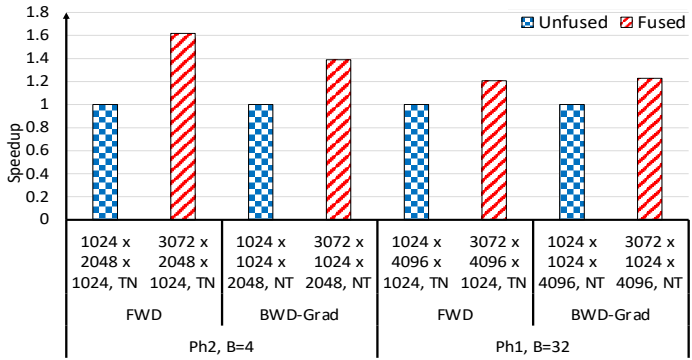

Figure 15: Impact of Fusing 3 Linear GEMMs (3F) in Attention layer, normalized to the non-fused (serial) versions.

These optimizations along with improvements in data layout have shown to speed up BERT training considerably [36].

### 5.2. Hardware Mechanisms

**Larger memory capacity**: A larger memory capacity on accelerators can significantly speedup training. A larger memory can enable training with a larger mini-batch size per device as it can hold more activations and gradients. It improves overall throughput through both fewer total training iterations, as well as larger, more efficient operations. A larger memory capacity also enables larger models to fit in a single device, and therefore, reduces or eliminates the communication overheads arising from model-parallel deployments.

**Larger on-chip (shared) memory**: Improving the accelerator's Last-Level Cache (LLC) to hold a layer's activation can help retain data between the producer and consumer layers or operations, avoiding redundant reads and writes to memory. This, however, can only improve performance in cases of consecutive producer-consumer layers. In the case of LAMB optimizer, for example, all the gradients are read after the entire backpropagation completes. Thus, there is minimal temporal locality in data. Moreover, the LAMB optimizer reads 4X as much data as the model size (about 4 GB), again with poor temporal locality, making it harder for such operations to benefit from a larger on-chip memory.

**Improved network bandwidth**: Improving network bandwidth is critical to efficiently scale out BERT training. As shown in Figure 12, the volume of data communicated increases with increasing number of devices in model-parallel training thus limiting the benefits of scaling. A higher network bandwidth can reduce these communication overheads thus improving the scaling efficiency at high device count.

**GEMM Accelerators**: A dynamically configurable accelerator to process the diverse sets of GEMMs. We have seen in Section 3.2.2 that GEMMs corresponding to the different sub-layers in BERT have different characteristics. While some GEMMs, such as those in the FC layer, are larger and compute-bound, others, such as those of the attention-heads, are small and memory-bound. Therefore, a GEMM's requirements can change as its size, shape or layout in memory change, making a single implementation and/or policy for a GEMM accelerator sub-optimal.

**In-Network Processing**: In-network processing involves acceleration of computations by adding compute capabilities to network switches. Such accelerators reduce the need for CPUs or GPUs to perform network-related operations, and eliminate the interference between computation and communication operations. These accelerators have especially shown considerable gains for collective operations such as All-Reduce [44], used heavily in both model and data-parallel distributed training (discussed in Section 4).

**Near-memory computing**: Near-memory computing (NMC) performs operations using specialized ALUs that are part of the main memory. Thus, NMC can be a faster and more energy-efficient alternative to accelerate memory-intensive operations as it does away with additional latency and energy to read and write data to and from memory [3, 43]. It further overcomes the capacity issue that on-chip memory face, as computations can span all the banks in memory, thereby providing the illusion of an extremely large on-chip cache. However, NMC requires changing the memory to add ALUs, and introduces new challenges such as programming NMCs, data alignment, and maintaining coherence between on-chip caches and the memory [13, 24].

## 6. Discussion

**Other accelerators**: Since we run BERT pre-training on a GPU, the runtime may differ on other GPUs or accelerators. However, the findings with respect to the manifestation, size, and compute characteristics are relevant to all accelerators. Thus, one can further (approximately) extrapolate our breakdowns for other devices by comparing the compute and memory bandwidth ratios between our GPU and the accelerator of interest.

**BERT Fine-tuning & Inference**: We focus on characterizing only the pre-training phase of BERT training. However, the key takeaways remain similar for both pre-training and fine-tuning since only the output layer of the network changes during the fine-tuning phase. The output layer of specific tasks, such as SQUAD (Q&A) [61], is simpler than tasks that BERT is pre-trained for, requiring fewer GEMMs and thus making the output layer a negligible component of SQUAD fine-tuning. Importantly, the transformer layers remain unchanged and still dominate the runtime, with similar breakdown compared to that of pre-training.

The execution profile of BERT inference will differ from pre-training since it does not require backpropagation or model parameter updates. Thus, the high-level breakdown of an iteration would not include LAMB update. However, since backpropagation has approximately 2X the number of operations as a forward pass, with near-similar properties, the breakdown of the transformer layers would largely remain similar to that of pre-training.

**Other NLP models**: As discussed in Section 1, although several other Transformer-based models have been proposed after BERT, we focus on BERT since it embodies several of the key trends that are important when designing and optimizing future accelerators for Transformer-based networks. Moreover, the overall breakdown and nature of operations would largely remain similar across these models.

## 7. Related Work

In recent years, there have been a large number of papers optimizing ML algorithms at a variety of levels of the compute stack. Broadly, the most relevant subsets of these papers can be broken into two groups: papers that characterize ML algorithms and papers that optimize Transformers.

**Characterizing ML Algorithms** [28, 54, 63, 78, 86, 88]: Prior work has characterized the behavior of ML workloads, especially the CNNs and RNNs in the MLPerf suite. Although the majority of this work has focused on characterizing inference [63, 78, 86], some of the work also characterizes CNN and RNN training [54, 88] or recommender systems [28]. However, unlike our work, prior work does not focus on Transformers, which are becoming increasingly important optimization targets for modern systems – especially the expensive pre-training phase we focus on in this work. Moreover, these works also do not provide detailed runtime breakdown amongst operations. Transformer-based networks such as BERT have received less attention. Works that include characterizations of Transformers either do not provide a detailed runtime breakdown amongst operations, only focus on Transformer's FC layers, or focus on inference rather than the expensive pre-training execution [29, 78, 80, 86].

**Optimizing Transfomers** [17, 29, 32]: Recent work has also examined how to design accelerators for Transformer-based networks. However, the relative lack of comprehensive characterization of Transformer-based models compared to the larger body of literature characterizing other ML algorithms has led prior work on designing accelerators for Transformer-based networks to overlook key characteristics of self-attention. For example, recent works design both efficient matrix-vector [29, 32] and matrix-matrix engines [17] to accelerate BERT training and/or inference [29, 32] even though the majority of the time, as shown in this work, BERT is not executing matrix-vector operations. Unlike in RNNs where tokens are processed one at a time, transformer layers calculate the relationship between *every* token pair in the input sequence, thus processing all the tokens of the input sequence in parallel. This leads to matrix, rather than vector, operations in transformer layers even if the mini-batch size is one (e.g., during inference). Furthermore, while self-attention requires multiplication of every input token vector with all the input vectors (a matrix) in the sequence, this operation is invoked for every token in the input sequence, thus requiring several such matrix-vector operations with a common matrix. Thus, these operations are usually performed as a single matrix-matrix operation to improve data reuse and exploit the inherent parallelism in the self-attention algorithm. Although some prior work acknowledges this property when comparing their accelerator against GPUs [29], it does not influence the design of the accelerator. This confusion in recent work about the use of matrix-vector operations in BERT underscores the necessity of our work, which focuses on understanding DNNs at an algorithmic level, before building efficient accelerators for them.

## 8. Conclusion

BERT has been a groundbreaking innovation in NLP providing extremely accurate language modeling capabilities. Its accuracy arises from its bi-directional transformer architecture, hundreds of millions of parameters, and its ability to train on enormous unlabeled datasets. Furthermore, its success has further inspired several popular models that are in use today. However, training these models is enormously expensive due to their large compute and memory requirements. Thus, they pose interesting challenges to system designers – challenges that must be met through deeper understanding of algorithmic behaviors as the waning of Moore's Law changes the virtuous synergy that has helped propel the transformative improvements of ML and NLP in recent years. Thus, in this paper we focus on BERT's most expensive component, pre-training, analyze its execution, and provide a detailed characterization of its compute requirements. Moreover, we further analyze how these characteristics change with evolving hyperparameters, training techniques including mixed precision, as well as in a distributed setting. By focusing on the architecture-agnostic components of BERT training, such as the description of the nature, size, and manifestation of its constituent operations, our analysis demystifies BERT's behavior and identifies holistic future acceleration opportunities.

## 9. Acknowledgements